\input{aipcheck}

\documentclass[
    ,final            
  ]
  {aipproc}

\layoutstyle{8x11single}

\begin{document}

\title{Orbifold GUT model with nine Higgs doublets}

\classification{12.10.Dm, 12.60.Jv}
\keywords      {SU(5), orbifold, higgs doublets}

\author{Biswajoy Brahmachari}{
  address={Department of Physics, Vidyasagar Evening College\\
39 Sankar Ghosh Lane, Kolkata 700006, India\\
and\\
Theory Division, KEK\\
 Tsukuba, Ibaraki 305-0801, Japan}
}

\begin{abstract}
We describe a non-supersymmetric orbifold GUT based on SU(5) symmetry.
It is a modification of Kawamura's 5-D orbifold GUT model. The difference 
lies in the choice of Higgs scalars as we have allowed only 5-plets of SU(5)
in the GUT scale. This variant was originally proposed by Brahmachari 
and Raychoudhuri. Proton decay problem and the doublet triplet splitting
problems are solved by extra dimensional mechanism. The unification
scale is around $5.0 \times 10^{13}$ GeVs. In low energy there are
nine Higgs doublets. One at the 100 GeV region and eight others
degenerate at around 1.4 TeV. It is an attractive non-supersymmetric 
extension of standard model with very rich collider physics phenomenology.
\end{abstract}

\maketitle


\section{Introduction}
The $SU(5)$ model\cite{su5} attempts to unify the strong, weak, and
electromagnetic interactions in the smallest simple group. It has many
attractive features which are well recognized, however it suffers
from a few major difficulties which are actually generic
to the idea of grand unification\cite{guts} itself.  

\begin{enumerate}
\item
Because quarks and leptons reside in unified multiplets and there are B- and
L-violating interactions, gauge boson exchanges can result in proton
decay\cite{pd}.  If these gauge bosons are appropriately heavy, the decay rate
will be very small. Their masses, in the usual formulation, are,
however, not arbitrary but rather determined by the scale where the
different gauge couplings unify.  The proton decay lifetime is
therefore a robust prediction of the model. No experimental signature
of proton decay\cite{pdexp} has been found yet and the model is
disfavoured. More complicated unification models involving several
intermediate mass-scales can evade this problem\cite{gaugeb}. 

\item 
The low energy Higgs doublet, responsible for electroweak breaking, is 
embedded in a {\bf 5} representation of $SU(5)$.  The other members of this 
multiplet are color triplet scalars which must have a mass near the 
unification scale and no such scalars have been observed at the 
electroweak scale. This leads to an unnatural mass splitting among the 
members of the same $SU(5)$ multiplet.  This is termed the double-triplet
splitting problem\cite{dt}.

\item 
The non-supersymmetric version does not have a natural dark matter
candidate\cite{dm}. Neutral 
members of $SU(5)$ particle spectram decay quickly.

\end{enumerate}
These unwelcome features of the $SU(5)$ model can be tackled in an
elegant way if unified $SU(5)$ symmetry exists in a 5-D world. Low
energy 4-D $SU(3)_c \times SU(2)_L \times U(1)_Y$ symmetry is
recovered when the extra dimension is compactified on a $S^1/(Z_2
\times Z^\prime_2)$ orbifold\cite{szz}. This situation is realized
when space-time is considered to be factorized into a product of 4D
Minkowski space-time $M^4$ and the orbifold $S^1/(Z_2 \times
Z^\prime_2)$. The coordinate system consists of
$x^\mu=(x^0,x^1,x^2,x^3)$ and $y=x^5$. There are two distinct 4-D
branes; one at $y = 0$ and another at $y =\pi R/2$. On the $S^1$, $y$=0 is
identified with $y = \pi R$ while
$y = \pm \pi R$/2 are identified with each other.. 

As in usual SU(5) model, fermions are put into ${\bf \overline{5}}$ plet
and the ${\bf 10}$ plet. We assume, as is common, that the fermions are fixed
in the 4-D brane at $y = 0$ whereas gauge bosons and the
scalars penetrate inside the bulk. The discrete $Z_2$ and
$Z^\prime_2$ symmetries, permit the expansion of any 5-D 
field $\phi$ in the following mode expansions according to 
whether they are even or odd fields.
\begin{eqnarray}
\phi_{++}(y) &=& \sqrt{4 \over 2^{\delta_{n,0}} \pi R} \sum_{n=0}^{\infty}
\phi^{(2n)}_{++}\cos{2 n y \over R};  ~~~~M_n = \frac{2n}{R} 
\label{eqn1} \\
\phi_{-+}(y) &=& \sqrt{4 \over \pi R} \sum_{n=0}^{\infty}
\phi^{(2n+1)}_{-+}\sin{(2 n +1) y \over R};  ~~~~M_n =
\frac{2n+1}{R}
\label{eqn2}\\
\phi_{+-}(y) &=& \sqrt{4 \over \pi R} \sum_{n=0}^{\infty}
\phi^{(2n+1)}_{+-}\cos{(2 n +1) y \over R};  ~~~~M_n = \frac{2n+1}{R}
\label{eqn3}\\
\phi_{--}(y) &=& \sqrt{4 \over \pi R} \sum_{n=0}^{\infty}
\phi^{(2n+2)}_{--}\sin{(2 n +2) y \over R};  ~~~~M_n = \frac{2n+2}{R} 
\label{eqn4}
\end{eqnarray}

\section{GUT symmetry breaking}
In this scheme ${\bf 24}$ Higgs has been excluded. Instead
of using Higgs mechanism, we use orbifold properties for breaking
$SU(5)$ symmetry. $A_{\mu}$ has + parity and $A_5$ has - parity under
$Z_2$. Further more (3,2) and ($\overline{3}$,2) components
of $A_\mu$ has negative $Z^\prime_2$ parity.
\begin{table}
\begin{tabular}{c c c}
\hline
{\bf Components} & {\bf $Z_2$} & {\bf $Z^\prime_2$} \cr
\hline
$A_\mu \rightarrow$ (1,1)+(1,3)+(8,1) & + & + \cr
$A_\mu  \rightarrow$ (3,2)+($\overline{3}$,2) & + & - \cr
$A_5 \rightarrow$ (1,1)+(1,3)+(8,1) & - & - \cr
$A_5  \rightarrow$ (3,2)+($\overline{3}$,2) & - & + \cr
\hline 
\end{tabular}
\caption{Parity assignments for different components of
5-D gauge field. These assignments were first 
given by Kawamura.}
\label{table1}
\end{table}
In other words $Z_2$ can distinguish between usual 4-D and
the extra fifth component of $A^M$ (5-D gauge field), 
whereas $Z^\prime_2$ can distinguish between SM gauge 
bosons and the extra SU(5) gauge bosons. Thus 
$Z^\prime_2$ assignments break SU(5) symmetry, because
a mass splitting among 24 gauge bosons are introduced. Out
of 24, 12 have mass-less modes but remaining 12 do not.
We choose $M_{GUT} = {1 \over R}$, where $R$ is
the radius of fifth dimension. Therefore GUT scale is
same as compactification scale.
\subsection{Proton decay problem}
It was noted by Dienes, Dudas and Gherghetta\cite{ddg}, that
if fermions are restricted to orbifold fixed points, all
$Z^\prime_2$ odd type wave functions, such as $X$ and $Y$
gauge bosons will vanish at orbifold fixed points. Thus there
is no coupling of $X$ and $Y$ gauge bosons to low energy
quarks and leptons, forbidding proton decay. Later on, 
Altarelli and Feruglio\cite{afhd} noted that the absence of
tree level amplitudes can provide an explanation of present
negative experimental results. They also noted that even-though
the idea of forbidding proton decay by a suitable discrete
symmetry is not new, its physical origin is clear in the
present context.
\begin{figure}
\includegraphics[height=.2\textheight]{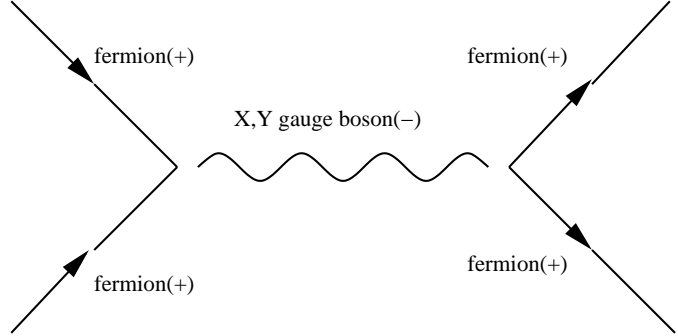}
\caption{$X$ and $Y$ gauge bosons cannot couple to fermions as
they have negative $Z_2^\prime$ parity. This makes the 
proton stable.}
\end{figure}
\subsection{Doublet triplet splitting}
Our model has only ${\bf 5}$ plets of Higgs fields. From 
Eqn. \ref{eqn1}- Eqn. \ref{eqn4}, we see that only $\phi_{++}$
has a mass-less mode. Therefore, once we include 
one 5-dimensional
${\bf 5}$ plet, one \underbar{four dimensional doublet} 
remains mass-less whereas the triplet has a mass of the 
order of the GUT scale. Similarly for `$n_5$' numbers of 5-dimensional
${\bf 5}$ plets, `$n_5$' numbers of \underbar{four dimensional doublets} 
remains mass-less at low energy. 
\begin{table}
$
\begin{array}{rcl}
\hline
SU(5) \supset && SU(3)_c \times SU(2)_L \times U(1)_Y  \nonumber \\
\hline
{\bf 5}    \supset && ( 1,2,1/2)_{\bf+~+} + (3,1,-1/3)_{\bf+~-} \nonumber\\
{\bf \overline{5}}    \supset && ( 1,2,-1/2)_{\bf+~+}
+ (\overline{3},1, 1/3)_{\bf+~-} \nonumber\\
{\bf 10}    \supset && ( 1,1,1)_{\bf+~+} + (\overline{3},1,-2/3)_{\bf+~+}
+(3,2,1/6)_{\bf+~-} \nonumber\\
{\bf 15}    \supset && ( 1,3,1)_{\bf+~+} + (3,2,1/6)_{\bf+~-}
+(6,1,-2/3)_{\bf+~+} \nonumber\\
{\bf 24}    \supset && ( 1,1,0)_{\bf+~+} + (1,3,0)_{\bf+~+}\nonumber\\
&& +(3,2,-5/6)_{\bf+~-} +(\overline{3},2,5/6)_{\bf+~-}+(8,1,0)_{\bf+~+} 
\nonumber\\
\hline
\end{array}
$
\caption{Decomposition of SU(5) representations up-to dimension
${\bf 24}$, where $Z_2$ and $Z^\prime_2$ are assigned in ${\bf 5}$ and
${\bf \overline{5}}$ only. Higher representations are obtained by 
group multiplication.}
\label{table2}
\end{table}
\subsection{Renormalization group equations}
\begin{figure}
  \includegraphics[height=.3\textheight]{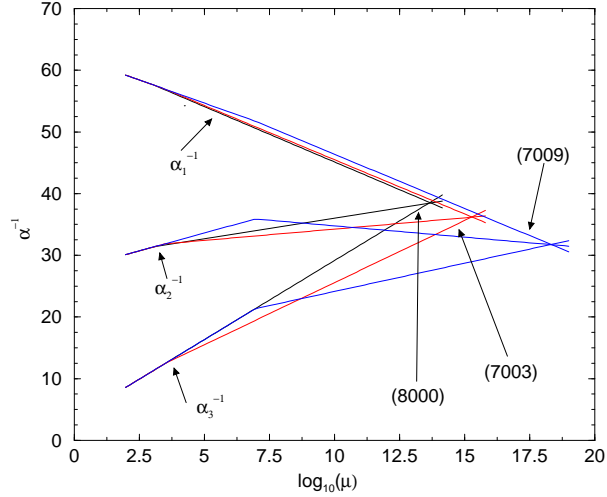}
  \caption{
Gauge unification in various models. Labels of cases
are $(n_5,n_{10},n_{15},n_{24})$. As a first approximation we
have used one intermediate scale which is given by the mass scale of 
extra scalars allowed by $S^1/Z_2 \times Z^\prime_2$ compactifications.
}
\label{fig2}
\end{figure}

Let us define new quantities $m_{k,l}= \ln({M_k/M_l})$ 
and $b^i_{k,l}$ coefficients range $M_k \leftrightarrow M_l$
\begin{eqnarray}
2 \pi \alpha^{-1}_i(M_Z) = 2 \pi \alpha^{-1}_{X}
+b^i_{X,I} M_{X,I} + b^i_{I,Z} M_{I,Z} 
\label{eqns}
\end{eqnarray}
Where $b$ coefficients are defined as,
\begin{eqnarray}
b^i_{X,I}=\pmatrix{41/10 \cr -19/6 \cr
-7 } + {n_5 \over 3} \pmatrix{3/10 \cr 1/2 \cr 0} 
\label{case1}
\end{eqnarray}
There are three equations and three unknowns. Solving for the
unknowns we get,
\begin{eqnarray}
&&\alpha^{-1}_{X}=38.53 \\
&&M_{I,Z}=26.98-194.75/n_5 \\
&&M_{X,I}=194.75/n_5 
\end{eqnarray}
Because $M_{I,Z} \ge 0$ we obtain $n_5 \ge 8$.
For the case of $n_5=8$ we get,
\begin{eqnarray}
M_I = 1.39~ {\rm TeV} ~~M_{X} =5.0 \times 10^{13}~ {\rm GeV} 
\end{eqnarray}
Therefore we see that this simple model predicts one doublet
at the $m_Z$ scale plus eight more doublets at around $1.4$ TeV.
Therefore this is a \underbar{nine Higgs doublet model}. 
Obviously, this model is very interesting from the point of
view of collider physics phenomenology. In Figure
\ref{fig2} we have plotted the unification scenario our
case has the label (8,0,0,0). More general cases can be 
found in the reference \cite{br} which
also discusses the present case in detail.
\section{Conclusions speculations and outlook}
There is proton decay non-observation problem in 
conventional GUTs. There is also doublet triplet splitting problem in
conventional GUTs. These two problems can be solved if SU(5) symmetry
exists in a 5-D world. The simplest model containing
only {\bf 5} plets has nine doublets $(1+8)$
at low energy. If we count the number of fermions in standard
model, there are $6$ quarks and $3$ leptons. Therefore the
total number of fermions are nine per generation. Perhaps
we have one doublet for each of them. It is an interesting puzzle.
We have compactified the fifth dimension on a 
${S^1/Z_2 \times Z^\prime_2}$ orbifold. These extra symmetries
can make the proton quite stable. This type of model being
non-supersymmetric, there is no R-parity violating proton
decay. There is no neutrino mass in this model. However
we can add heavy right handed neutrinos in the GUT scale.
This is a heavy right handed singlet fermion of mass $10^{13}$ GeV or so.
Then via see-saw mechanism one can produce a small majorana mass
of the order of
\begin{eqnarray}
m={m^2_D \over M_{GUT}} 
\end{eqnarray}
Here, we can see that, in our case $M_{GUT}=5 \times 10^{13}$. Therefore,
we can make an order of magnitude estimate of the neutrino mass,
\begin{eqnarray}
m = {(10^2)^2 \over 5 \times 10^{13}} = 0.2 \times 10^{-9} {\rm ~GeV} 
= 0.2 {\rm ~eV}. 
\end{eqnarray}
We should study Kaluza-Klein type dark matter in this
model. The lightest Kaluza-Klein state with (at-least one) negative 
parity could be stable
and its lifetime can be comparable to the age of the universe.
It can be seen that $Z_2$ parity is the same for all multiplets
of $scalars$. Only the $Z^\prime_2$ changes. Therefore is $Z_2$ 
redundant ? We have
kept it here even though it is of no use $for~scalars$. 
This is because if we
want to construct the supersymmetric version of the theory, we may have
to use it. One can study $b-\tau$ unification in this model. The simplest
way is to couple only one doublet to Fermions. However all nine doublets
contribute to gauge coupling unification. 
At last let us comment on the 5-D Planck scale where quantum gravity becomes
non-negligible,
\begin{eqnarray}
M^{5d}_{planck}=M^{2/3}_P \times M^{1/3}_c
\end{eqnarray}
For our case, $M_P=10^{19}$ GeV and $M_c=M_{GUT}=10^{13}$ GeV, Therefore 
the 5-D Planck scale is at
\begin{eqnarray}
M^{5d}_{planck}=10^{16.87}~~{\rm GeV}
\end{eqnarray} 
We see that the Planck scale is higher than unification
scale as expected. This result follows from our requirement that
GUT scale and 5-D compactification scale should be  one and the same.

\begin{theacknowledgments}
This work was supported by UGC, New Delhi, India, under the
grant number F.PSU-075/05-06
\end{theacknowledgments}



\bibliographystyle{aipproc}   


\begin{thebibliography}{9}

\bibitem{su5}
H. Georgi, S.L. Glashow, Phys. Rev. Lett. {\bf 32}438 (1974);
H. Georgi, H. R. Quinn, S. Weinberg, Phys. Rev. Lett. {\bf 33},
451 (1974).


\bibitem{guts}
J. C. Pati, A. Salam, Phys. Rev. {\bf D10}, 275 (1974);
Jogesh C. Pati, A. Salam, Phys. Rev. {\bf D8}, 1240 (1973);
R. N. Mohapatra, J. C. Pati, Phys. Rev. {\bf D11}, 2558 (1975);
F. Gursey, P. Ramond, P. Sikivie, Phys. Lett. {\bf B60}, 177 (1976)

\bibitem{pd}
A.J. Buras, J. R. Ellis, M.K. Gaillard, D.V. Nanopoulos,
Nucl. Phys. {\bf B135}, 66 (1978); C. Jarlskog,
Phys. Lett. {\bf B82}, 401 (1979);
C. Jarlskog, F.J. Yndurain,
Nucl. Phys. {\bf B149}, 29 (1979); F. Wilczek, A. Zee, 
Phys. Rev. Lett. {\bf 43}, 1571 (1979); M. Machacek,
Nucl. Phys. {\bf B159}, 37 (1979)

\bibitem{pdexp}
SuperKamiokande Collaboration (Y. Hayato et al.),
Phys. Rev. Lett. {\bf 83} 1529 (1999); 
Super-Kamiokande Collaboration (M. Shiozawa
et al.), Phys. Rev. Lett. {\bf 81} 3319 (1998);
W. Gajewski et al., Phys.Rev. {\bf D42}, 2974 (1990);
Soudan-2 Collaboration (W.W.M. Allison et al.), Phys. Lett. 
{\bf B427} 217 (1998).


\bibitem{dm}G. Bertone, D. Hooper, J. Silk, Phys. Rept. {\bf 405} 279 (2005). 
e-Print: hep-ph/0404175 


\bibitem{gaugeb}P. H. Frampton, B. Hoon Lee, Phys. Rev. Lett. {\bf 64}
619 (1990); P. H. Frampton, T. W. Kephart, Phys. Rev. {\bf D42}, 
3892 (1990);  B. Brahmachari, U. Sarkar, R. B. Mann, T. G. Steele, 
Phys. Rev. {\bf D45}, 2467 (1992); 
B. Brahmachari, U. Sarkar, Phys. Lett. {\bf B303}, 260 (1993);
P. B. Pal, Phys. Rev. {\bf D45}, 2566 (1992); B. Brahmachari, 
Phys. Rev. {\bf D48}, 1266 (1993)

\bibitem{dt}
H. Georgi, Phys. Lett. {\bf 108B}, 283 (1982); B. Grinstein,
Nucl. Phys. {\bf B206}, 387 (1982); A. Masiero, D.V.
Nanopoulos, K. Tamvakis, T. Yanagida, Phys. Lett. {\bf B115}
380 (1982); S. Dimopoulos, F. Wilczek, NSF-ITP-82-07 (unpublished);
A. Sen, Phys. Rev. Lett. {\bf 55}, 33 (1985);

\bibitem{ddg}
K. R. Dienes, E. Dudas, T. Gherghetta, 
Phys. Lett. {\bf B436} 55 (1998); Nucl. Phys. {\bf B537} 47 (1999). 


\bibitem{szz}
Y. Kawamura, Prog. Theor. Phys. {\bf 103}, 613 (2000);
Y. Kawamura, Prog. Theor. Phys. {\bf 105}, 691 (2001);
A. Hebecker, J. March-Russell, Nucl. Phys. {\bf B613}, 3 (2001);
J. A. Bagger, F. Feruglio, F. Zwirner, Phys. Rev. Lett. {\bf 88}
101601 (2002); Y. Nomura, D. R. Smith, N. Weiner,
Nucl. Phys. {\bf B613}, 147 (2001); L. J. Hall, Y. Nomura, Phys. Rev. 
{\bf D64}, 055003 (2001).

\bibitem{afhd}
G. Altarelli, F. Feruglio, Phys. Lett. {\bf B511}, 257 (2001); 
A. B. Kobakhidze, Phys. Lett. {\bf B514}, 131 (2001).


\bibitem{br}
B. Brahmachari, A. Raychaudhuri, J. Phys. {\bf G29} B5 (2003). 

\end{thebibliography}




\end{document}